\documentclass[b5paper,twoside]{jpconf}  
\usepackage{graphicx}

\begin{document}

\title[The slow acoustic mode in the flare]{The slow magnetoacoustic mode in the flaring loop}

\author[V.Reznikova and K.Shibasaki]{Veronika Reznikova \& Kiyoto~Shibasaki}

\address{Nobeyama Solar Radio Observatory/NAOJ, Nagano 384-1305, Japan}
\ead{reznik@nro.nao.ac.jp; shibasaki@nro.nao.ac.jp} 
\begin{abstract}
We studied long duration flare observed with Nobeyama Radioheliograph at frequencies 17 and 34 GHz and with Ramaty High Energy Solar Spectroscopic Imager at 25-50 keV. We found that microwave and hard X-ray emission variation contain well-pronounced periodicity with the oscillation period growing from 2.5 to 5 min. Analysis of the loop length and plasma temperature evolution during the flare allowed to interpret the quasi-periodic pulsations in terms of the second standing harmonics of the slow magnetoacoustic mode. This mode can be generated by the initial impulsive energy release and work as a trigger for the repeated energy releases.
\end{abstract}
\section{Introduction}
Quasi-periodic pulsation (QPP) with periods from tens of seconds to tens of minutes are often observed in solar flare light curves in radio and X-ray bands (see \cite{Nakariakov2006}, for a review). These long period coronal oscillations are believed to be associated with magnetohydrodynamic (MHD) waves or/and to be generated by magnetic reconnection. Therefore, these oscillations are of fundamental importance, because they can provide information about mechanisms responsible for the energy release, its triggering, and processes operating in them. 

\section{Observations}\label{Obs}
\begin{figure}[ht]
\begin{minipage}{15pc}
\includegraphics[width=13pc]{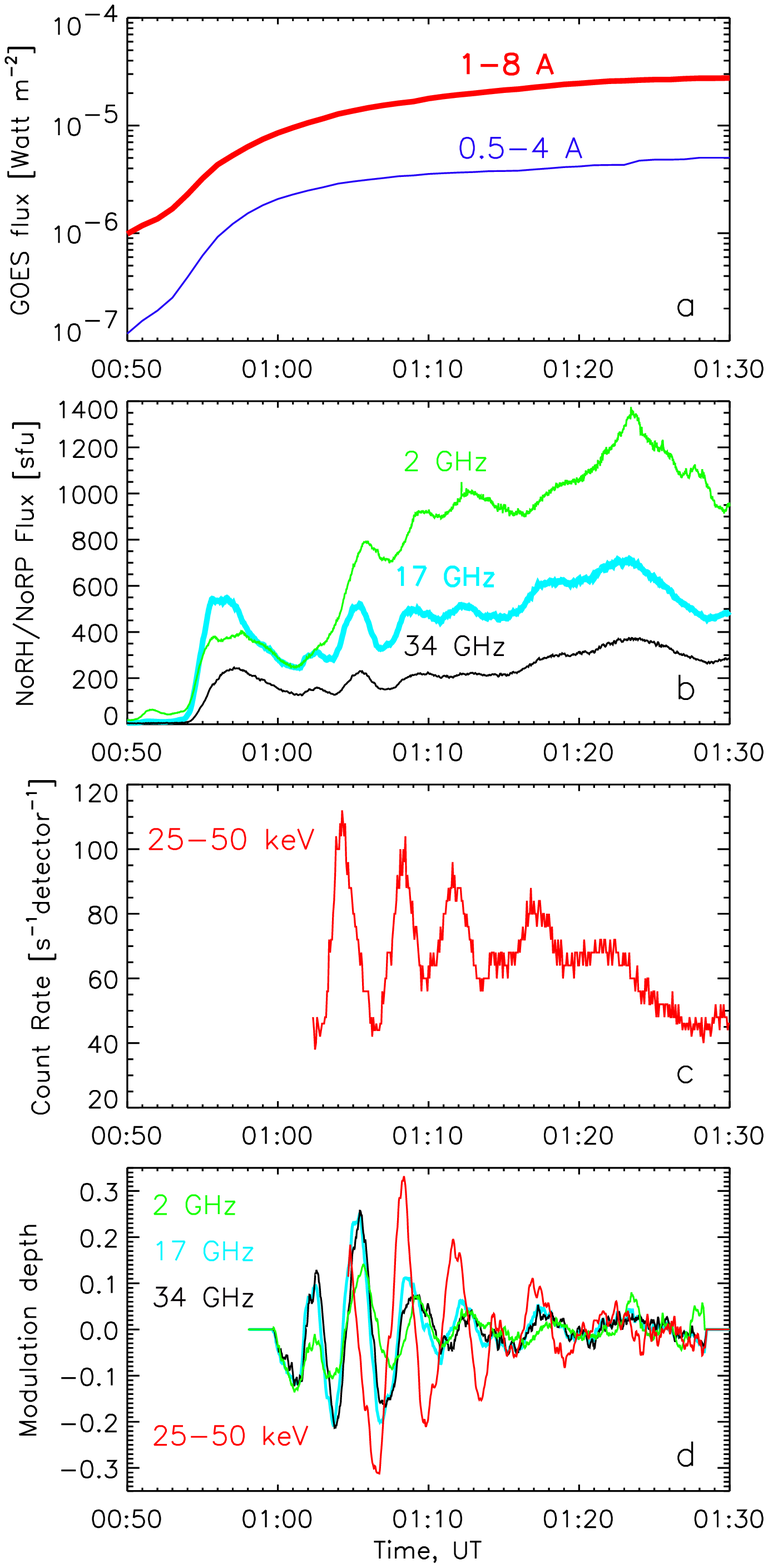}
\caption{\label{fig1}Light curves of the flare. (a) Soft X-ray in the GOES 1-8 \AA\ (thick) and 0.5-4 \AA\ (thin) channels; (b) NoRP flux at 2 GHz (green) and NoRH fluxes at 17 GHz (blue), and 34 GHz (black); (c) hard X-ray count rate measured in RHESSI 25-50 keV channel; (d) modulation depth of emission at frequencies 2 GHz (green), 17 GHz (blue), 34 GHz (black) and energy 25-50 keV (red).}
\end{minipage}\hspace{1pc}%
\begin{minipage}{15pc}
\includegraphics[width=15pc]{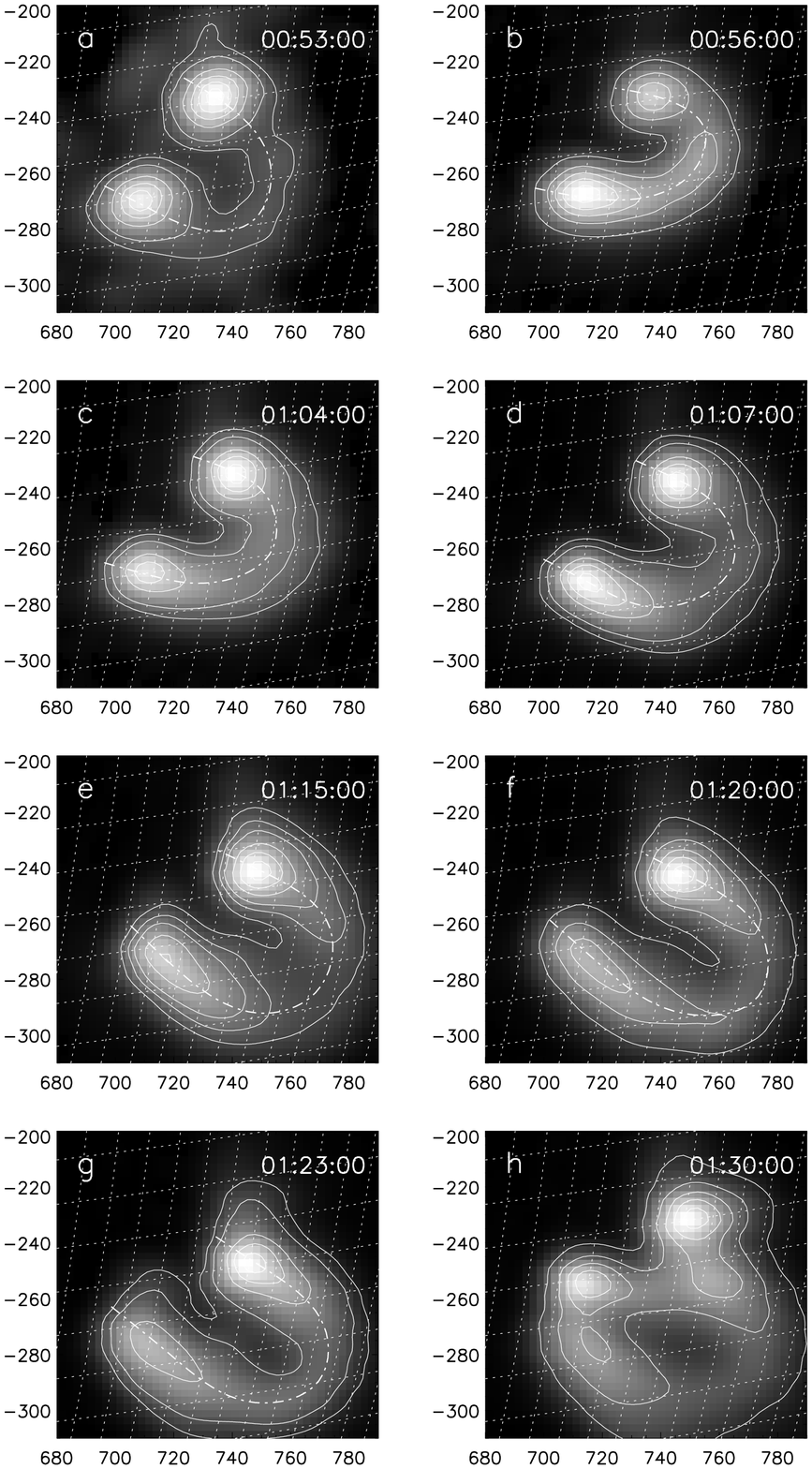}
\caption{\label{fig2}NoRH images of the flare source at 17 GHz (I). Contours correspond to 0.1, 0.3, 0.5, 0.7, 0.9 levels of maximum $T_{b}$. The brightness of filling increases with the increasing $T_{b}$. The dot-dashed lines show visible flaring loop axes. The units of the X- and Y-axis are arcsec from the disk center.}
\end{minipage} 
\end{figure}

The intense solar flare occurred 2005 August 22 at 00:54 UT with heliographic coordinates S12$^\circ$, W49$^\circ$ and was connected with the active region NOAA 10798. GOES fluxes in channels 1-8 \AA\ (thick line) and 0.5-4 \AA\ (thin line) are shown in Fig. \ref{fig1}a. The flare was of the M2.6 class on the GOES scale and was related to the long duration events. The corresponding microwave burst was observed by the Nobeyama Radioheliograph (NoRH) at two frequencies, 17 and 34 GHz, and by Nobeyama Radiopolarimeter (NoRP). Figure \ref{fig1}b shows the NoRP time profile at 2 GHz and the NoRH time profiles of the total fluxes integrated over the partial images, the size of $313''\times313''$, which includes the whole AR. They display multiple emission peaks at all frequencies.

The Ramaty High Energy Solar Spectroscopic Imager (RHESSI) was in the shadow of the Earth until 01:02 UT, therefore, observations of the X-ray burst started only after the main peak on the NoRH/NoRP time profiles. The RHESSI count rate in the channel 25-50 keV is shown in Fig. \ref{fig1}c. The time profiles of hard X-rays clearly display damped quasi-periodical pulsations. 
In Figure \ref{fig1}d we show the modulation depth of emissions observed by three different instruments, NoRP (2 GHz), NoRH (17, 34 GHz), and RHESSI (25-50 keV). The modulation depth of the signal is calculated as 
$\Delta F/F = (F(t)-F_0)/F_0.$ The slowly varying mean signal $F_0$ is obtained by 200 s smoothing of the signal.
Radio emission at all three frequencies shows pulsations with an in-phase behavior and a modulation depth of up to 20\%. The hard X-ray emission exhibits a higher modulation depth of up to 30\%. 

Figure \ref{fig2} shows images of the flare loop at 17 GHz (Stokes I) at different moments of the burst. 
The entire loop structure was well seen during all the burst, but its length and orientation changed with time. At about 01:23 UT significant changes in the flare morphology began (Fig. \ref{fig2}g) and subsequently a system consisting of two loops appeared (Fig. \ref{fig2}h). The similar evolution was observed in intensity at 34 GHz.
\begin{figure}[h]
\begin{minipage}{15.5pc}
\includegraphics[width=15pc]{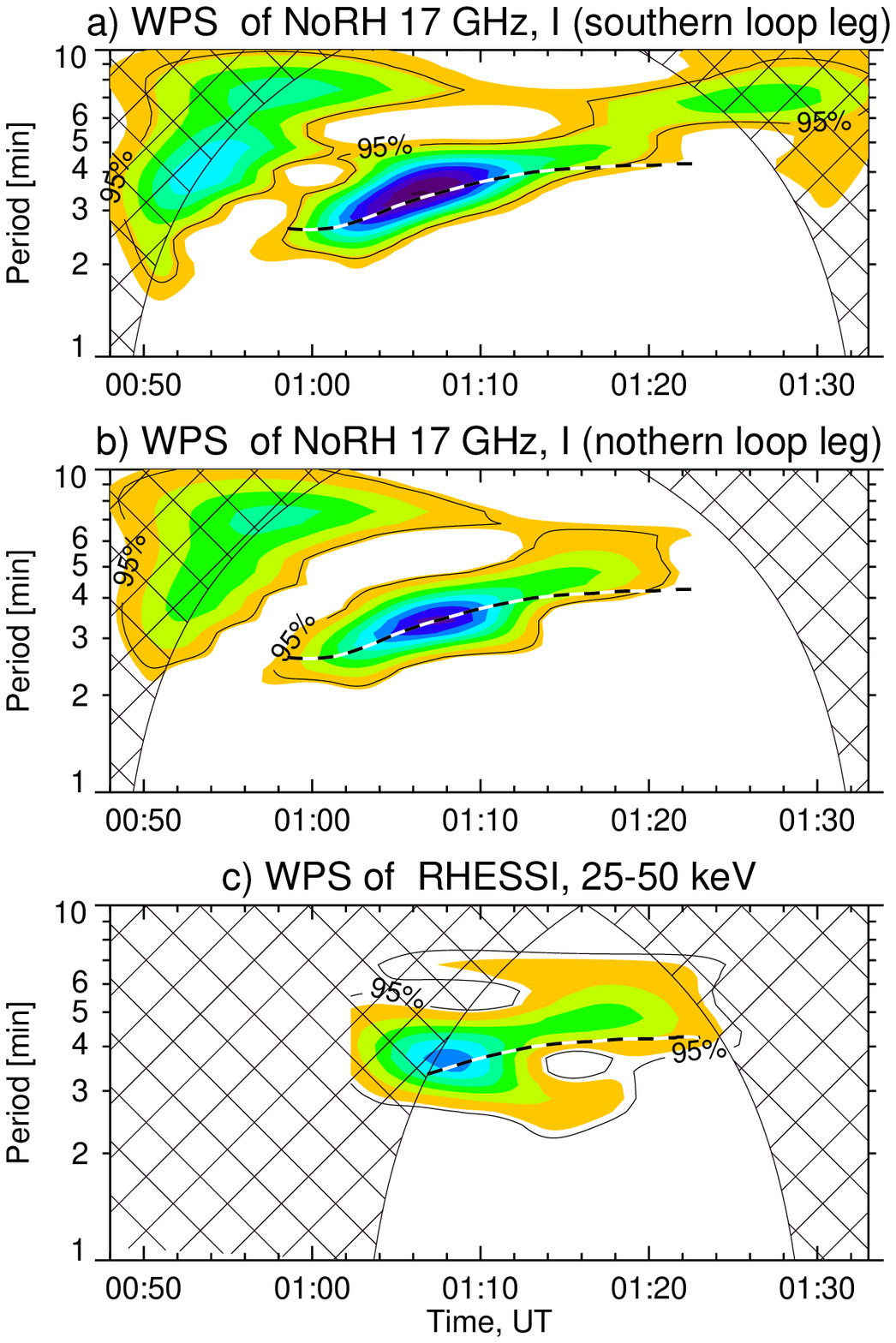}
\caption{\label{fig3} Wavelet power spectra for 17 GHz signals from (a) southern loop leg, (b) northern loop leg and (c) RHESSI hard X-ray signal in 25-50 keV channel integrated over the source.} 
\end{minipage}\hspace{1.2pc}%
\begin{minipage}{15pc}
\includegraphics[width=12.5pc]{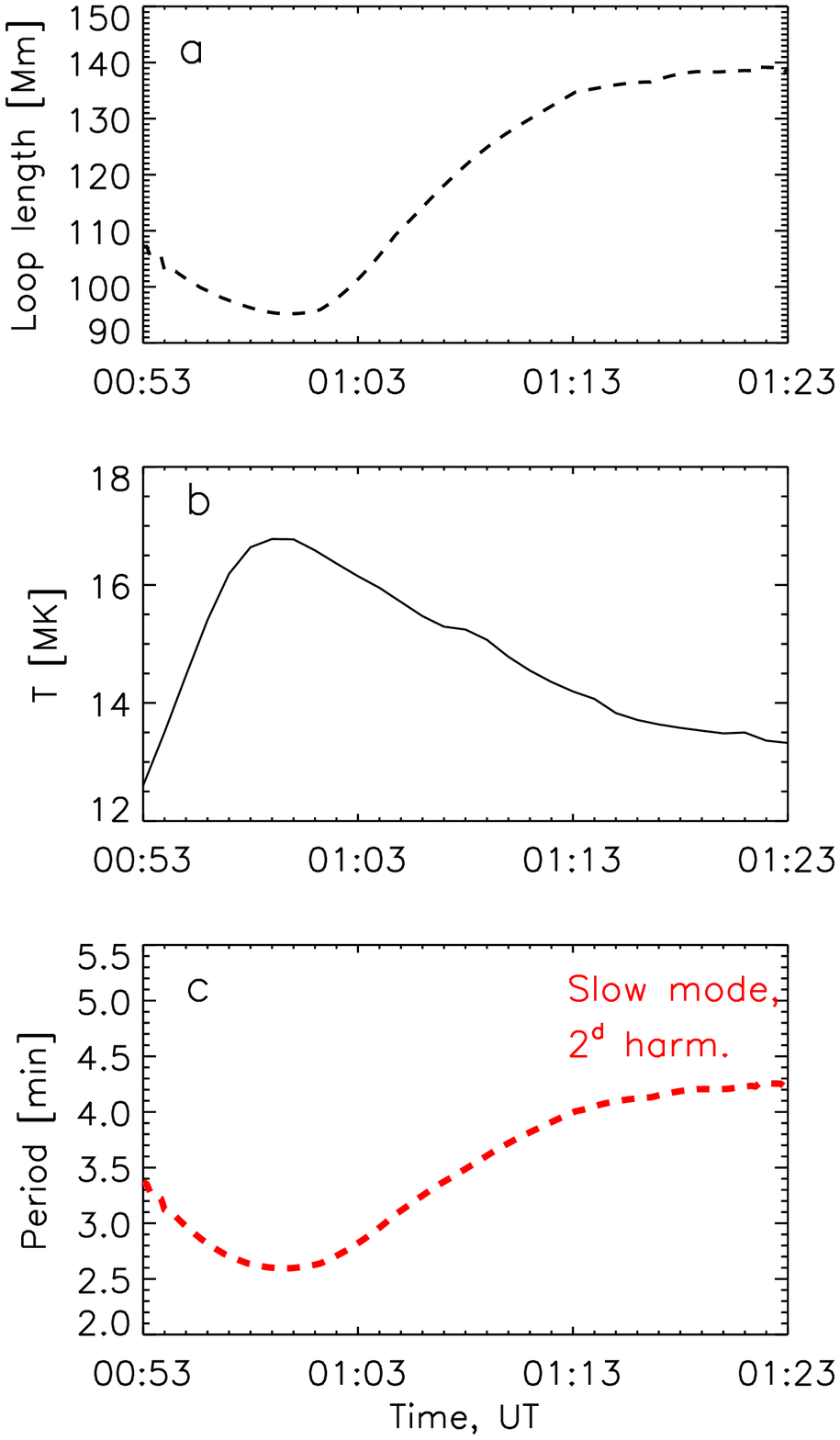}
\caption{\label{fig4} Time profiles of (a) the loop length measured along deprojected loop axis, (b) plasma temperature, and (c) period of the slow magnetoacoustic wave, the second harmonics.} 
\end{minipage}
\end{figure}
\section{Spectral analisys and interpretation}\label{discus}
The microwave and hard X-ray signals were processed by means of the wavelet analysis. The wavelet power spectra are shown in Fig.~\ref{fig3}. They contain a periodic component growing from 2.5-3 min to 4-5 min between 01:00 UT and 01:23 UT. This component appeares immediately after the first emission peak and agrees well with the period of the second harmonics of the slow magnetoacoustic wave shown by the dashed line. This period was calculated using formula from \cite{Nakariakov2006}:
\begin{equation}
P/s = k \times (L/Mm)/\sqrt{(T/MK)}, \label{period}
\end{equation}
where $k\approx6.7$ for the second harmonics, $L$ is the loop length (in Mm), and $T$ is the average temperature in the loop (in MK). 

To estimate the loop length we used a series of 17 GHz images. The apparent loop length is measured along the visible loop axis shown in Fig.\ref{fig2}. To obtain values of the deprojected loop length we conducted a loop axis reconstruction using a geometrical technique \cite{Loughhead1983}. Estimates of the plasma temperature were made from the analysis of the integral soft X-ray fluxes in the two channels 1-8~\AA\ and 0.5-4~\AA\ recorded by GOES.

The time profiles of a reconstructed (or deprojected) loop lengths smoothed over 200 s are shown in Fig. \ref{fig4}a. We used the deprojected loop length for estimating the slow magnetoacoustic wave period. Figure \ref{fig4}b shows the evolution of the plasma temperature, time resolution is 1 min. It is growing during the first microwave peak up to $T = 16.7~$MK and then gradually decreases. Time profiles of estimated period is shown in Fig.~\ref{fig4}c for the second harmonics. The relative error of the estimate $\Delta P/P=0.07$. 

\section{Discussion}\label{dis}
All analyzed wavelet spectra contain a periodic component that grows with time in the course of the flare. This component agrees well with the calculated period of the second harmonics of the slow magnetoacoustic wave. It appears immediately after the first emission peak and disappears when the magnetic configuration undergoes strong changes: the flare transfers to a different loop system. 

It was demonstrated by Nakariakov \cite{Nakariakov2004} that the second standing acoustic harmonics may appear as a natural response of the loop to an impulsive energy deposition located near the loop top. Moreover, Tsiklauri \cite{Tsiklauri2004} showed that only the second harmonics is excited, regardless of the spatial position of heat deposition. They modeled the evolution of a coronal loop in response to an impulsive heating. It was shown that first, a sudden energy release results in a rapid local increase in the pressure gradient and loop density because of the chromospheric evaporation. The strong impulsively generated pressure disturbance propagates as a slow magnetosonic wave along the loop legs and is reflected at the footpoints finally as a standing wave. 

We estimated the background plasma density inside the loop \cite{Reznikova2011} and excluded the possibility of the microwave and hard X-ray emission modulation caused by a density variation in the wave. The high modulation depth of both wavelengths emission indicates that QPP is likely to be a manifestation of periodical magnetic reconnection. Therefore, we suppose that the slow wave excited by the initial energy release may work as a trigger for the repeated energy releases.
\ack
The work was partly supported by RFBR grant No. 11-02-91175. The author is grateful to Prof. V. Nakariakov and Dr. V. Melnikov for useful discussions. Wavelet software was provided by C. Torrence and G. Compo, and is available at http://paos.colorado.edu/research/wavelets/.

\end{document}